\documentclass[twocolumn,aps,prl,showpacs,preprintnumbers]{revtex4}%
\usepackage{mathrsfs}
\usepackage{amsmath}
\usepackage{amsfonts}
\usepackage{amssymb}
\usepackage{amsthm}
\usepackage{graphicx}
\usepackage{color}
\usepackage{graphics}
\usepackage{epsfig}
\usepackage{subfigure}%
\providecommand{\U}[1]{\protect\rule{.1in}{.1in}}
\begin{document}
\title{Domain wall propagation through spin wave emission}
\author{X.S. Wang$^{1}$, P. Yan$^{2}$, Y.H. Shen$^{1}$, G.E.W. Bauer$^{3,2}$ and X.R.
Wang$^{4,1}$}
\email{[Corresponding author:]phxwan@ust.hk}
\affiliation{$^{1}$Physics Department, The Hong Kong University of Science and Technology,
Clear Water Bay, Kowloon, Hong Kong}
\affiliation{$^{2}$Kavli Institute of NanoScience, Delft University of Technology, Delft,
The Netherlands}
\affiliation{$^{3}$Institute for Materials Research, Tohoku University, Sendai 980-8577, Japan}
\affiliation{$^{4}$School of Physics, Wuhan University, Wuhan, P. R. China}

\begin{abstract}
We theoretically study field-induced domain wall (DW) motion in an
electrically insulating ferromagnet with hard- and easy-axis anisotropies. 
DWs can propagate along a dissipationless wire through spin wave emission 
locked into the known soliton velocity at low fields. In the presence of 
damping, the mode appears before the Walker breakdown field for strong 
out-of-plane magnetic anisotropy, and the usual Walker rigid-body 
propagation mode becomes unstable when the field is between the 
maximal-DW-speed field and Walker breakdown field.

\end{abstract}

\maketitle


Magnetic domain-wall (DW) propagation in nanowires has attracted attention
because of the academic interest of a unique non-linear system
\cite{Walker,Cowburn,Erskine,Parkin} and potential applications in data
storage and logic devices \cite{Parkin,Review,NEC}. The field-driven DW
dynamics is governed by the Landau-Lifshitz-Gilbert (LLG) equation
\cite{Walker}, which has analytical solutions in limiting cases
\cite{Walker,xrw}, such as the soliton solution \cite{Braun} in the absence of
both dissipations and external magnetic fields. The interplay between spin
waves (SWs) and DWs has also received attention, including DW propagation
driven by externally generated SWs \cite{kim,Seo,magnon} and SW generation by
a moving DW \cite{Wieser,Yanming}. Our understanding of the field-induced DW
motion is nevertheless far from complete. According to conventional wisdom DWs
move under a static magnetic field only in the presence of energy dissipation
\cite{Walker,Wang}. Numerical evidence against this view therefore came as a
surprise \cite{Wieser}.

\begin{figure}[ptbh]
\begin{center}
\includegraphics[width=8.0cm]{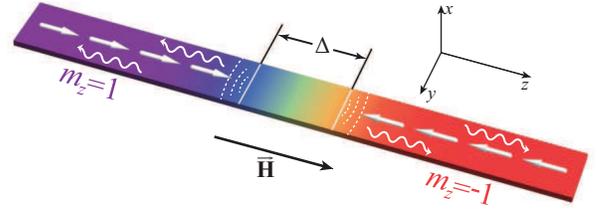}
\end{center}
\caption{(color online) Schematic of a head-to-head DW of width $\Delta$ 
in a magnetic nanowire. $\vec{\mathbf{H}}$ is an external field along 
wire-axis defined as $z$-direction. DW breathing and other types of 
periodic DW texture deformations emits spin waves, denoted by the 
wavy lines with arrows.} 
\label{fig1}%
\end{figure}

We report here a physical picture for the SW emission-induced domain wall
motion for a head-to-head DW in a magnetic nanowire with easy axis along the
wire ($z$-direction) as shown in Fig. \ref{fig1}. Let $K_{\parallel}$ and
$K_{\perp}$ be anisotropy coefficients of the easy and hard axis (along the
$x$-direction), respectively. An external field along the wire rotates the DW
out of the $yz$-plane. The DW structure thereby experiences an internal field
in the $x$-direction twisting the DW plane and generating a non-uniform
internal field along the wire. This field causes periodic deformations of the
DW structure, such as \textquotedblleft breathing\textquotedblright%
\ \cite{Walker} by which the entire DW precesses around the wire axis while
its width shrinks and expands periodically.
The local modulation of the magnetization texture generates SWs (wavy lines
with arrows in Fig. 1) that radiate away from the DW center. The energy needed
to generate the SWs has to come from the Zeeman energy \cite{Wang} that is
released by propagating the DW. The DW velocity of a dissipationless
ferromagnet in the steady state may then be expected to be proportional to the
SW emission rate. 

In this Letter, we numerically solve the LLG equation, initially without
damping in order to confirm the above mentioned relation between spin wave
emission and DW propagation. Depending on $K_{\perp}$ and the magnetic field,
breathing or more complicated periodic transformations of the DW emit spin
waves. The DW propagation at low fields tends to lock into a particular
soliton mode in which the energy dissipation rate due to the SW emission is
balanced by the Zeeman energy gain. We predict robust spin wave emission that
persists in the presence of Gilbert damping and renders the usual Walker
rigid-body propagation mode unstable in region below the Walker breakdown field.

The LLG equation reads
\begin{equation}
\frac{\partial\mathbf{m}}{\partial t}=-\mathbf{m}\times\mathbf{h}%
_{\mathrm{eff}}+\alpha\mathbf{m}\times\frac{\partial\mathbf{m}}{\partial t},
\label{LLG}%
\end{equation}
where $\mathbf{m}$ is the unit direction of the local magnetization
$\mathbf{M}=\mathbf{m}M_{s}$ with saturation magnetization $M_{s}$ and
$\alpha$ is the Gilbert damping constant. The effective magnetic field of our
biaxial wire (see Fig. 1) is $\mathbf{h}_{\mathrm{eff}}=K_{\parallel}m_{z}%
\hat{z}-K_{\perp}m_{x}\hat{x}+{A}\partial^{2}\mathbf{m/}\partial{z^{2}}%
+H\hat{z},$ consisting of internal and external fields in the unit of $M_{s}$.
$A$ is the exchange constant. Time, length and energy density are measured in
units of $\left(  \gamma M_{s}\right)  ^{-1}$ with gyromagnetic ratio $\gamma
$, the DW width at equilibrium $\Delta_{0}=\pi\sqrt{A/K_{\parallel}}$, and
$\mu_{0}M_{s}^{2},$ respectively. We chose parameters of the electric
insulator Yttrium Iron Garnet (YIG) with \cite{magnon,para} $A=3.84\times
{10^{-12}}$ $%
\operatorname{J}%
/%
\operatorname{m}%
$, $K_{\parallel}=2\times10^{3}$ $%
\operatorname{J}%
/%
\operatorname{m}%
^{3}$, $\gamma=35.1$ $%
\operatorname{kHz}%
/\left(
\operatorname{A}%
/%
\operatorname{m}%
\right)  $ and $M_{s}=1.94\times10^{5}$ $%
\operatorname{A}%
/%
\operatorname{m}%
$, and the corresponding time and length units are $1.46\times10^{-10}$ $%
\operatorname{s}%
$ and $1.38\times10^{-7}$ $%
\operatorname{m}%
$, respectively. $K_{\perp}$ and $\alpha$ are treated as adjustable parameters
depending on the sample shape and microscopic order. We solve Eq. \eqref{LLG}
by a numerically stable method \cite{Algorithm}. The mesh size is chosen to be
$0.009$, corresponding to the YIG lattice constant ($1.24$ $%
\operatorname{nm}%
$). \begin{figure}[ptb]
\begin{center}
\includegraphics[width=8.5cm]{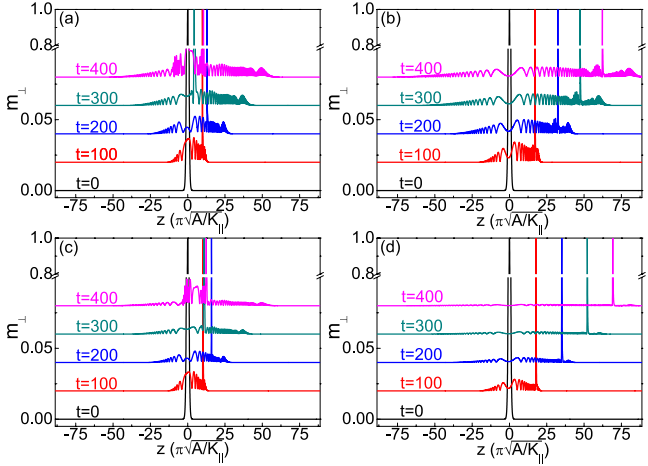}
\end{center}
\caption{(color online) Snapshots of $m_{\perp}=\sqrt{m_{x}^{2}+m_{y}^{2}}$ 
for $H=0.01$ at various times $t$ for (a) $K_{\perp}=4$ and $\alpha=0$; 
(b) $K_{\perp}=10$ and $\alpha=0$; (c) $K_{\perp}=4$ and $\alpha=0.001$; 
(d) $K_{\perp}=10$ and $\alpha=0.001$.} 
\label{fig2}%
\end{figure}

\begin{figure}[ptbh]
\begin{center}
\includegraphics[width=8.5cm]{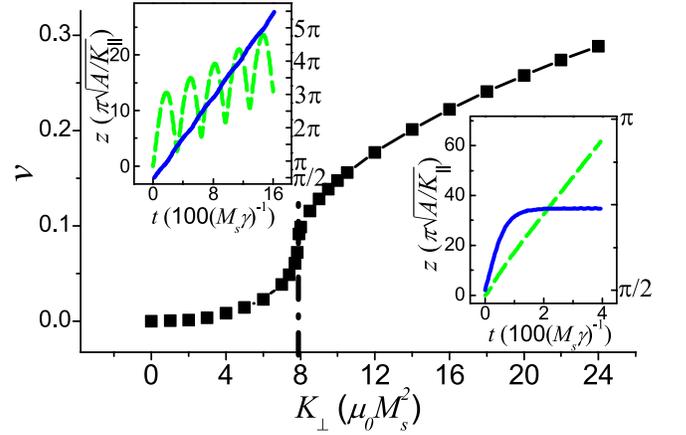}
\end{center}
\caption{(color online) $K_\perp$-dependence of the DW velocity $v$ (in units 
of $\pi\gamma M_s\sqrt{A/K_\parallel}$) for fixed $H=0.01$ and $\alpha=0$. 
Insets: Time dependence of the DW center position (dashed line) and the 
azimuthal angle $\phi$ (solid line) of magnetization at the DW center for 
$K_{\perp}=4$ (left) and $K_{\perp}=10$ (right).} 
\label{vk2}%
\end{figure}

To prove that a DW under an external field indeed emits spin waves, we plot
the snapshots of the distribution of $m_{\perp}\equiv\sqrt{m_{x}^{2}+m_{y}%
^{2}}$ for $K_{\perp}=4$ and $10$ (in units of $\mu_{0}M_{s}^{2}$), $\alpha
=0$, and $H=0.01$ (on at $t=0$) at various times in Figs. 2a and 2b. At $t=0$,
right before the external field is switched on, $m_{\perp}$ follows the Walker
DW profile \cite{Walker}, the DW center is located at the center ($z=0$) of a
wire with length $180\Delta_{0}$, while the DW magnetization lies in
$yz$-plane. Curves are offset for better visibility. As time proceeds ($t>0$),
SWs (wavy features) are emitted into both directions, while the DW center
(peak) moves simultaneously along the field slower than the SWs. The
velocities $v$ for a fixed magnetic field increase monotonically with
$K_{\perp}$ (Fig. \ref{vk2}). It does not depend sensitively on small
$K_{\perp}\left(  <4\right)  $, but grows rapidly when $K_{\perp}$ is close to
$8$, and becomes an almost linear function of field for $K_{\perp}>12$. From
the time-dependence of the position (dashed lines) of DW center and its
azimuthal angle (solid lines) shown in the insets of Fig. \ref{vk2} for a
small and large $K_{\perp},$ we trace the periodic DW deformations in the
different field regions. We note that such large $K_{\perp}$ value may be
realized in YIG samples subject to mechanical strains \cite{book}.

At a small $K_{\perp}=4$ (left inset), $\mathbf{m}$ at the DW center rotates
around the wire, while the center position oscillates back and forth but also
moves slowly along the applied field. This oscillatory motion is synchronized
with the breathing, \textit{i.e.}, the periodic energy absorbing and releasing
in the form of periodic oscillation of the DW width $\Delta=\pi\sqrt
{A/(K_{\parallel}+K_{\perp}\cos^{2}\phi)}$ \cite{Walker}, where $\phi$ is the
tilt angle of the DW plane (with equilibrium value $\pi/2$). This breathing
excites spin waves as shown in Fig. 2a. Low velocities corresponds to weak SW
emission. At a large $K_{\perp}=10$ (right inset), the azimuthal angle of the
DW center approaches a fixed value and the DW center position moves at a
constant velocity since $\phi$ and the DW energy are almost constant. In this
case, $\mathbf{m}$ still rotates around the wire axis while the DW center
propagates along the wire with a fixed $\phi$. The large $K_{\perp}$ twists
the DW plane into a chiral screw-like structure that changes shape
periodically during the magnetization precession while the DW center
\textquotedblleft drills\textquotedblright\ forward.
This drilling mode is much more efficient in emitting SWs than the breathing
mode, leading to a relatively high DW propagating speed. For YIG parameters
the SW velocity exceeds that of the DW, therefore, in contrast to Ref.
\cite{Wieser}, we observe bow as well as stern SW excitations.

\begin{figure}[ptbh]
\begin{center}
\includegraphics[width=8.5cm]{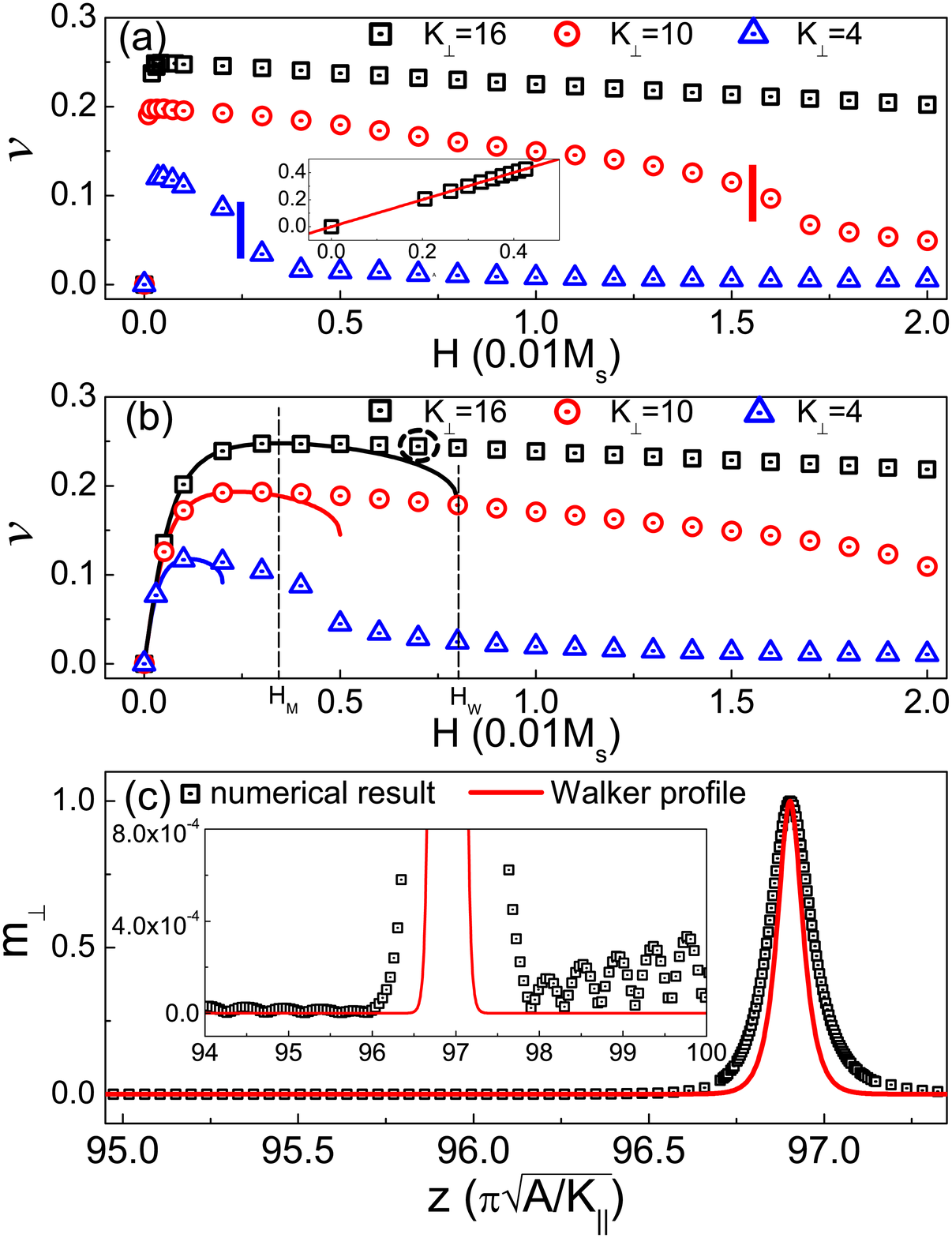}
\end{center}
\caption{(color online) (a)(b) Field dependence of DW velocity for various 
$K_\perp$. Field and velocity are in units of $0.01M_{s}$ and $\pi\sqrt{A/
K_\parallel}\gamma M_s,$ respectively. $A$, $K_\parallel$ and $M_s$ are 
YIG parameters. (a) $\alpha=0$. Inset: symbols display $-\sin2\phi$ 
\textit{vs.} $v\sqrt{(K_{\parallel}+K_{\perp}\cos^{2}\phi)/A}/K_{\perp}$, 
where $v$ is the soliton velocity. The line is the linear relationship 
from soliton theory. (b) $\alpha=0.001$. The solid curves are the 
corresponding Walker solutions. (c) Snapshot of $m_{\perp}$ for $H=0.007$ 
and $K_\perp=16$ at $t=400$ (for the datapoint indicated by a circle in (b)). 
The numerical results deviates from the Walker profile that is clearly 
narrower. Inset: Expanded view of $m_{\perp}$ close to the DW  center. 
The relative deviation from the Walker profile (solid curve) is pronounced 
and spin wave emission is conspicuous. } 
\label{v_h}%
\end{figure}

Figure \ref{v_h}a displays the steady state DW velocity in a dissipationless
wire with $K_{\perp}=4,\ 10,$ and $16$ with parameters otherwise identical to
those of Fig. 3. As a function of field it increases abruptly for small
values, reaches a maximal value, and decreases again. When $K_{\perp}$ is
reduced from 16 to 4 the DW changes from a drilling to the breathing motion,
resulting in a significant drop of DW velocity. The decrease of DW velocity
with field should not be interpreted as suppression of SW emission or damping
of DW propagation by spin wave emission \cite{Wieser,Suhl}. The Zeeman energy
released by the DW motion at a rate $2HMv$ \cite{Wang} should be equal to the
energy rate carried away by the SWs. Therefore, provided that the latter
increases sub-linearly with $H$ ($\propto H^{\beta}$ and $\beta<1$), the DW
propagation speed must decrease with field. The initial rapid rise of the DW
velocity at small fields is related to the soliton solution $\ln\tan
(\theta/2)=\pi(z-vt)/\Delta$ of the LLG Eq. (1) with soliton velocity
$v=-\sqrt{A/(K_{\parallel}+K_{\perp}\cos^{2}\phi)}K_{\perp}\sin2\phi$ for
$\alpha=0$ and $H=0$ \cite{Braun}, where $\theta$ is the polar angle of
$\mathbf{m}$. This can be seen from the plot of $-\sin2\phi$ for the saturated
$\phi$ \textit{vs.} $v\sqrt{(K_{\parallel}+K_{\perp}\cos^{2}\phi)/A}/K_{\perp
}$ (symbols) for $H\in\lbrack0,0.0009]$ in the inset of Fig. \ref{v_h}a for
$K_{\perp}=16$. The numerical data agree precisely with the soliton velocity
formula (solid line). Since conventional solitons do not satisfy the LLG
equation in the presence of an external field we unearthed here a hybridized
mode of solitons and spin waves. In contrast to the zero field case, there is
only one particular soliton mode in which the SW emission is balanced by the
Zeeman energy change, \textit{viz}. the moving DW. This holds as long as the
field is smaller than the value at which $-\sin2\phi$ reaches its maximum
value 1. Beyond that field, the soliton mode becomes instable since the SW
emission rate cannot keep up with the released Zeeman energy and other
propagation modes have to take over. This soliton instability point is
emphasized by vertical bars in the data points for $K_{\perp}=4$ and
$K_{\perp}=10$, indicating the threshold fields below which the soliton
formula holds. For $K_{\perp}=16$ this value is out of the range of Fig.
\ref{v_h}a.

For YIG parameters and $K_{\perp}=10$, the DW velocity at $H=0.01\left(
\sim24\text{ }\mathrm{Oe}\right)  $, is about $140$ $%
\operatorname{m}%
/%
\operatorname{s}%
$. For comparison, the corresponding DW velocity by Walker rigid body
propagation for the same parameters and $\alpha=0.05$ is $57$ $%
\operatorname{m}%
/%
\operatorname{s}%
$ \cite{Walker}. We may conclude that the DW velocity in low dissipation
magnetic insulators is of the same order of magnitude as in highly dissipative
ferromagnetic metals. 

DW propagation through spin wave emission exists in any magnetic wire with
transverse magnetic anisotropy irrespective of the Gilbert damping. Figures
\ref{fig2}c and \ref{fig2}d look very similar to Figs. \ref{fig2}a and
\ref{fig2}b in spite of the finite damping $\alpha=0.001$. Naturally, in the
presence of damping the SWs can propagate only over finite distances, which
explains why they have been overlooked in most experimental and numerical
studies. In Fig. 2b (or 2a) and Fig. 2d (or 2c), the DW velocity is higher in
the presence of a small non-zero damping, since damping dissipates an energy
on the top of the SW emission and the DW velocity is proportional to the
energy dissipation rate \cite{Wang}. Here we find a mixed DW propagation mode
that profits from both Gilbert damping and spin wave emission. Figure
\ref{v_h}b is the field dependence of the DW velocity for $\alpha=0.001$ and
various $K_{\perp}$, while other model parameters are unmodified from Fig.
\ref{v_h}a. Except at very small fields, Fig. \ref{v_h}b is similar with Fig.
\ref{v_h}a. The numerically obtained velocities (symbols) in Fig. \ref{v_h}b
agree well with Walker's rigid-body propagation (solid curves)
\cite{Walker,Yanming} below some maximum field $H_M\simeq 2(K_\parallel
/K_\perp)^{0.25} H_W $ for $K_\perp >> K_\parallel$, where 
$H_W=\alpha K_{\perp}/2$ \cite{Walker} is the Walker breakdown field.  
However, deviations are obvious for fields between $H_{M}$ and $H_W$ 
(where the solid lines end). 
This implies that Walker rigid-body propagation is not stable for 
$H\in\lbrack H_{M},H_{W}]$ with respect to the spin wave emission mode.

In order to prove numerically that the Walker solution is only stable for
$H<H_{M}$, we solve the LLG Eq. \eqref{LLG} starting from an initial
magnetization configuration that deviates slightly from the rigid-body
propagation mode. The magnetization indeed converges to the Walker profile for
$H<H_{M}$. However, this changes when we consider $H=0.007$ for $K_{\perp}=16$
(indicated by dashed circle in Fig. \ref{v_h}b) with all other parameters the
same. According to Walker theory \cite{Walker}, $H_{M}=0.00344$ and
$H_{W}=0.008$ (indicated by dashed lines in Fig. \ref{v_h}b). The symbols in
Fig. \ref{v_h}c give a snapshot of $m_{\perp}$ for $t=400$, at which the
transients die out and the DW center propagates to about $z=97$. The
distribution deviates significantly from the Walker profile (solid curve). 
The spin wave emission is clearly observed in the expanded tail in the 
inset of Fig. \ref{v_h}c. Exact analytic solutions do not make 
numerical methods obsolete.

There are several corollaries of the DW propagation mode by spin wave
emission. The emitted spin waves from one DW, for example, can mediate an
attractive force on the nearby DW, since a DW moves against passing spin waves
by spin transfer \cite{magnon}. This causes crosstalk in wires with more than
one domain walls with consequence for the \textquotedblleft race
track\textquotedblright\ memory \cite{Parkin}. This DW-DW attractive force has
a finite range governed by the Gilbert damping and is sensitive to material
parameters and geometry. The increase of the effective damping by SW emission
\cite{Suhl} is not restricted to DWs but will appear in any time-dependent
magnetization texture including magnetic vortices and cannot be captured by
the Gilbert phenomenology. On the other hand, our results possibly open
alternatives to manipulate and control the effective damping in magnetic nanostructures.

In conclusion, we prove that a DW in a wire with a finite transverse magnetic
anisotropy undergoes a periodic transformation under an external magnetic
field that excites SWs. The energy carried away must be compensated by the
Zeeman energy that is released by DW propagation along the wire. The DW
propagation adopts one particular soliton velocity at low fields. This SW
assisted DW propagation can be attributed to the SW emission generated by DW
breathing and drilling modes. In the presence of damping it competes with and
appears before the Walker breakdown. Also, the spin wave emission will mediate
a dynamic attractive force between DWs with a range controlled by the Gilbert damping.

This work is supported by Hong Kong UGC/CERG grants (\# 604109,
HKUST17/CRF/08, and RPC11SC05), the FOM foundation, DFG Priority Program
SpinCat, and EG-STREP MACALO.

\end{document}